# Confinement of Light in Disordered Photonic Lattices: A New Platform for Waveguidance

**Somnath Ghosh[1] and Bishnu P. Pal[*]**
[1]*Institute of Radio Physics and Electronics, University of Calcutta, Kolkata 700009, India*
[*]*Mahindra École Centrale, Bahadurpally, Hyderabad 500043, India;*

**Abstract:** A right amount of disorder in the form of refractive index variation has been introduced to achieve transverse localization of light 1D semi-infinite photonic lattices. Presence of longitudinally-invariant transverse disorder opens-up a new waveguiding mechanism.

## 1. Introduction

Guided wave optics and photonics essentially rely on the phenomenon of total internal reflection (TIR) though other mechanisms could also form its physical functional principle e.g. continuous refraction in a graded refractive index medium or photonic bandgap-mediated guidance (PBG) in certain photonic crystal geometries.[1,2]. A core of refractive index higher than the cladding that surrounds it is imperative for TIR guidance, whereas a long range order in the dielectric distribution is mandatory for PBG. Even though huge progress has already been achieved in guided wave optics, researchers in recent years have started exploring newer mechanism (s) for waveguiding. Breaking the symmetry of the system, length scale (upto tens of nanometer) of the refractive index distribution, as well as dimension of the system dictates the mode of novel guiding mechanisms. Readily available matured state-of-the-art fabrication techniques are the key driving force behind these studies. Waveguides based on surface plasmons is an example for such guidance [3].

In this context we investigate an alternative light guiding mechanism that exploits the phenomena of Anderson localization (AL) of a wave [4]. In a 1D photonic lattice we introduce disorder in the form of refractive index variation (longitudinal direction mapped onto time) and achieved the confinement of light along the transverse direction of light propagation [5]. The dependence of the localized state on the shape of an optical beam when it is coupled into an evanescently coupled, disordered, waveguide lattice has been studied. Our study revealed several underlying interesting features of light confinement to a localized state in such a medium of finite length. We have studied the nature of input light coupling to different localized modes of a disordered medium under different input excitation conditions. We also explicitly show that beyond the point of localization, light indeed propagates without any diffractive spread in the transverse direction in a disordered lattice, a feature that mimics waveguide-like propagation [6]. The presence of local nonlinearity of the lattice favors the formation of localized state. We had performed an experiment using ultrafast laser inscription (ULI) technique to demonstrate our numerical findings [7]. Our results should be useful for manipulating the flow of light in novel discrete photonic superstructures.

## 2. Photonic lattice with disorder

We consider an evanescently coupled waveguide array (*as photonic lattice*) consisting of a large number ($N$) of unit cells, and in which all the waveguides spaced equally apart are buried inside a medium of constant refractive index $n_0$ [5,7]. The overall structure is homogeneous in the longitudinal (z) direction along which the incident optical beam is assumed to propagate. The change in refractive index $\Delta n(x)$ (over the uniform background of $n_0$) due to disorder in this 1D waveguide lattice is assumed to be of the form

$$\Delta n(x) = \Delta n_p (H(x) + C\delta(x)) \qquad (1)$$

here $C$ is a dimensionless constant, whose value governs the strength of disorder; the periodic function $H(x)$ takes the value 1 inside the higher-index regions and is zero elsewhere; $\Delta n(x)$ consists of a deterministic periodic part $\Delta n_p$ of spatial period $\Lambda$ and a spatially periodic random component $\delta$ (uniformly distributed over a specified range varying from 0 to 1). This particular choice of randomly perturbed refractive index in the high index as well as low index layers enables us to model the *diagonal* and *off-diagonal* disorders to study the localization of light. Wave (1+1) $D$ propagation through the lattice is governed by the standard scalar Helmholtz equation, which under paraxial approximation can be written as

$$i\frac{\partial A}{\partial z} + \frac{1}{2k}\left(\frac{\partial^2 A}{\partial x^2}\right) + \frac{k}{n_0}\Delta n(x) A = 0 \qquad (2)$$

where $A(x,z)$ is amplitude of an input CW optical beam having its electric field as

$$E(x,z,t) = \text{Re}[A(x,z)e^{i(kz-wt)}]; \quad k = n_0\omega/c \qquad (3)$$

We solve Eq. (2) numerically through the scalar beam propagation method, which we have implemented in Matlab®. In our optimized 10 *mm* long 1D lattice geometry of 150 identical evanescently coupled waveguides, we consider the high index ($n_2$) regions of width ($d_2$) 7 μm, which are separated by low index regions ($n_1$) equal distances ($d_1$) of 7 μm (as shown in Fig. 1a). The value of $\Delta n_p$ was chosen to be 0.001 (*negligible bandgap*) over the background material of refractive index ($n_0$) 1.454.

Moreover, in an optical geometry having PBG, the phenomenon of light localization due to the inherent Bragg scattering by the ordered structure is fundamentally different from the effect of AL in the presence of a disorder. Accordingly, we have chosen a particular distribution of the transverse refractive index of the perfectly ordered lattice such that it spawns a prominent bandgap and introduced right amount of disorder. It has been shown that this could be an additional degree of freedom to control the guiding mechanism due to disorder.

### 3. Localization and light confinement

We consider a Gaussian beam (FWHM 10 μm) at the wavelength of 980 nm to be incident on the unit cell located around the central region of the lattice (*irrespective of the index of the local lattice unit*) and then set different values for the disorder parameter *C*. We estimate the effective widths ($\omega_{eff}$) of the output beam along the lattice length. The results are shown in Fig. 1b from which it can be seen that above a threshold value of *C* the beam opens up a waveguide-like channel through the lattice and get localized at the output end. Four different input beam shapes namely Gaussian, parabolic, hyperbolic secant and exponential respectively shows a very similar behavior implying that the transverse localization phenomena is nearly independent of the input beam shapes (true eigenstate of the system). Thus, the disordered photonic lattice structure behaves like a guiding geometry for the incident light, a phenomenon which has also recently been demonstrated in a cylindrical geometry, which led to the concept of an *Anderson localized fiber* (self-averaging must be strong enough to ensure that the localization can be observed in one particular sample rather than through ensemble averaging) [6,8,9]. In Fig. 1c we depict a typical localized state at the end of a 15 mm long 60% disordered optical waveguide lattice for a Gaussian input beam. The linearly decaying tail on a logarithmic scale carries the signature of localization. Incorporating the statistical nature of the disorder (over 100 ensembles of same disorder level) we have estimated the *localization length* [5]. In [6] it has been shown that the effective propagating beam diameter of localized light is comparable to that of a typical index-guiding optical fiber. Our investigation also revealed that formation of the localized state is nearly independent of the input position of the incident beam though precise shape of the localized eigenstate depends on the location of the input beam.

To further study the impact of the medium's nonlinearity upon this localization phenomenon, we have investigated the influence of focusing/defocusing types of Kerr nonlinearity through the parameter: $n_2|A|^2/\Delta n_p$ where $n_2$ is Kerr coefficient of the media. The interplay of focusing nonlinearity and level of disorder is shown in Fig. 1d, which revealed that a focusing nonlinearity favors localization of light in a disordered lattice. We have also observed a delocalization effect when a defocusing nonlinearity has been chosen. Thus one could achieve a nearly diffraction free propagation of light in a discrete disordered photonic structure in the presence of nonlinearity.

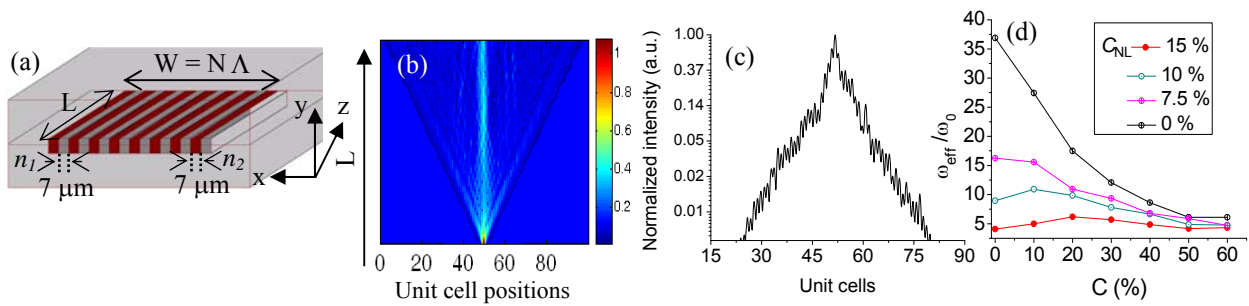

Figure 1. a) Schematic of the chosen photonic lattice with no disorder, b) Transition to a localized state in a 20% disordered lattice; c) Anderson localized state on a logarithmic scale at the end of a lattice with *C* = 0.6; (d) variation of the effective beam width (averaged over 100 realizations) with *C* for different values of focusing nonlinearity ($C_{NL}$).

## 4. Ultrafast laser inscripted photonic lattices in glass and localization

A lattice consisting of an array of a large number of waveguides was fabricated using a Fianium® Femtopower Yb-doped fiber laser emitting 350 fs pulses at a wavelength of 1064 nm. The repetition rate of the laser was set to 500 kHz. The pulses were focused onto the Erbium doped Bismuthate glass through an aspheric lens of 0.4 NA [see Fig. 2 (a)]. Changing the writing speed allows fine control of the net amount of laser energy transferred to the substrate, which essentially dictates the net modification of the refractive index. A range of arrays with varying $C$'s as scaling factor was fabricated [microscopic view in Fig 2 (b)], using the same pseudo-random array pattern for each. Moreover, in order to address the important issue of the statistical nature of the phenomenon in a finite-sized lattice, we fabricated lattices of different realizations for a particular $C$. Light from a 980 nm fiber laser at a relatively low power level (to avoid the possibility of any nonlinear effects which may arise due to the inherent nonlinearity of the sample) was launched into the fabricated waveguide lattice around its central region through an intermediate microscope objective. A CCD camera was employed to record the intensity distribution from the image of the lattice output. It can be seen from Fig. 2 (c) representing the measured output intensity distribution from a disordered lattice having $C = 0.6$, indicates clearly the signature of a localized state of light. Figure 2(d) shows a measured ensemble averaged localized intensity profile from waveguide lattices having C = 0.4.

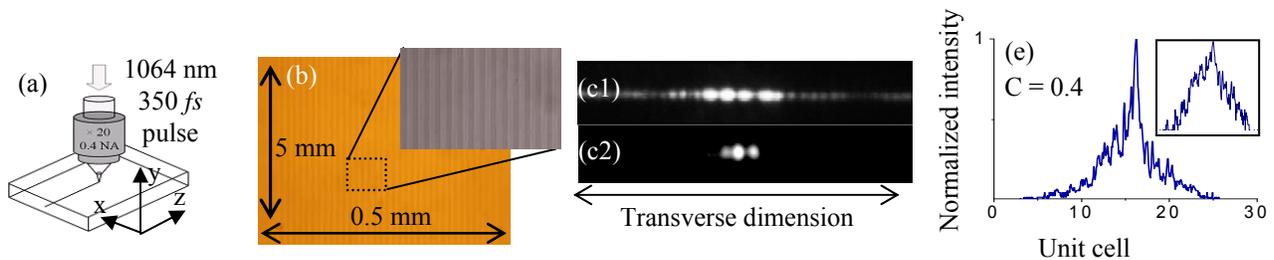

Fgure 2. (a) Schematic of the femto-second laser inscription technique. (b) Microscopic view of a ULI lattice of coupled waveguides, the inset shows the zoomed version of the marked area; (c1) Measured output intensity distribution from the lattice under ballistic mode of propagation in arbitrary units. (c2) Measured output intensity distribution corresponding to a localized state for a disordered lattice having $C = 0.6$. (d) Measured ensemble averaged output intensity profiles from waveguide lattices having $C = 0.4$, which corresponds to threshold for realization of localized light; the inset represents the localized intensity profile on a semi-log scale.

This talk would include how Anderson localized fiber/ waveguiding geometries are becoming increasingly attractive for technological applications in bio-imaging (in spite of the presence of a high level of disorder in the imaging system), short-haul fiber optical communications etc [8]. We also discuss the huge potential of these guides in future device applications including sensing and Anderson localized random nano-lasers.

This work relates to Department of the Navy Grant N62909-10-1-7141 issued by Office of Naval Research Global, which partially supported this work. The United States Government has royalty-free license throughout the world in all copyrightable material contained herein. SG acknowledges the financial support by Department of Science and Technology, India as a INSPIRE Faculty Fellow [IFA-12; PH-13].

## 5. References

[1] B. P. Pal (ed), Guided Wave Optical Components and Devices: Basics, Technology and Applications, (Academic Press?Elesevier, Burlington, 2006).
[2] P. Russell, Photonic Crystal Fibers, Science **299**, 358 (2003).
[3] S. A. Maier and H. A. Atwater, Plasmonics: Localization and Guiding of Electromagnetic Energy in Metal/Dielectric Structures, J. Appl. Phys. **98**, 011101 (2005).
[4] P. W. Anderson, "Absence of diffusion in certain random lattices," Phys. Rev. **109**, 1492-1505 (1958).
[5] S. Ghosh**,** G. P. Agrawal, B. P. Pal and R. K. Varshney, "Localization of light in evanescently coupled disordered waveguide lattices: Dependence on the input beam profile," Opt. Commun. **284**, 201-206, (2011).
[6] S. Karbasi et al, "Observation of transverse Anderson localization in an optical fiber", Opt. Lett. **37**, 2304 (2012).
[7] S. Ghosh, N. D. Psaila, R. R. Thomson, B. P. Pal, R. K. Varshney, and A. K. Kar, "Ultra-fast laser inscribed waveguide lattice in glass for direct observation of transverse localization of light", App. Phys. Lett. **100**, 101102 (2012).
[8] S. Karbasi et al, "Image transport through a disordered optical fibre mediated by transverse Anderson localization". *Nature Comm.* **5**, 3362 (2014).
[9] R. G. S. El-Dardiry, S. Faez, and A. Lagendijk, "Snapshots of Anderson localization beyond the ensemble average", Phys. Rev. B. **86**, 125132 (2012).